# Applying Spiking Neural Nets to Noise Shaping

Christian Mayr
Infineon Technologies AG
TU Dresden
Lehrstuhl hochparallele VLSI-Systeme und Neuromikroelektronik

René Schüffny
TU Dresden
Lehrstuhl hochparallele VLSI-Systeme und Neuromikroelektronik

*Abstract*-In recent years, there has been an increased focus on the mechanics of information transmission in spiking neural networks. Especially the Noise Shaping properties of these networks and their similarity to Delta-Sigma Modulators has received a lot of attention. However, very little of the research done in this area has focused on the effect the weights in these networks have on the Noise Shaping properties and on post-processing of the network output signal. This paper concerns itself with the various modes of network operation and beneficial as well as detrimental effects which the systematic generation of network weights can effect. Also, a method for post-processing of the spiking output signal is introduced, bringing the output signal more in line with conventional Delta-Sigma Modulators. Relevancy of this research to industrial application of neural nets as building blocks of oversampled A/D converters is shown. Also, further points of contention are listed, which must be thoroughly researched to add to the above mentioned applicability of spiking neural nets.

## I. INTRODUCTION

In biology, neural nets are able to transmit signals very rapidly and faithfully, even if these signals have spectral components well above the mean firing rate of the net, which is inconsistent with the notion of a rate coding [3]. Also, biological evidence indicates that in the majority of cases information is transmitted by groups of neurons rather than a single one. This has led researchers to investigate possible group-based information transmission mechanisms such as relative phase coding, correlation coding, etc. Recordings from neuron groups as well as simulation results point to the use of Noise Shaping as a transmission mechanism [2,3,4]. Noise Shaping is well known from Delta-Sigma-Modulator theory [1].

In its basic form, a Delta-Sigma-Modulator (DSM) consists of one or more cascaded integrators, a 1-bit quantizer (ADC) and a feedback loop, with the whole loop operating at a much higher frequency than the signal to be converted. In addition to the input signal, there is a high noise level present in the bit stream at the output because of the low number of quantizer levels in the ADC. Because this quantizer noise is entered into the feedback loop at a different point from the signal, it is subjected to a different filter function. The coefficients in a DSM are chosen in such a way that input signals in its base band are transmitted undisturbed, while the quantizer noise is filtered out of the base band towards higher frequencies. This behaviour is called Noise Shaping.

The makeup of the usual technical model of a spiking neuron, i.e. the Eckhorn neuron, is very close to a first-order DSM, containing an integrator, thresholded spike-generator (comparable to the ADC), and a feedback network. A number of researchers have shown that biological as well as technical spiking neural networks exhibit the same Noise Shaping behaviour as a DSM [2,3,4].

The main advantage of neural nets compared to DSMs stems from the difference in modes for achieving a high quantization rate, i.e. the oversampling ratio (OSR). While a conventional DSM contains only a single loop, which must operate at a frequency 10 to 100 times greater than the highest signal frequency, a neural net contains numerous modulators operating in parallel, with the combined pulse rate of the whole net constituting the oversampled signal (Fig. 1). These nets can even transmit, with high fidelity, signals well above the mean firing rate of a single neuron [2,3].

A possible application of these neural nets (including the pulse post processing proposed in this paper) in analog-digital converters and structural comparison with conventional DSMs is given in Fig. 1.

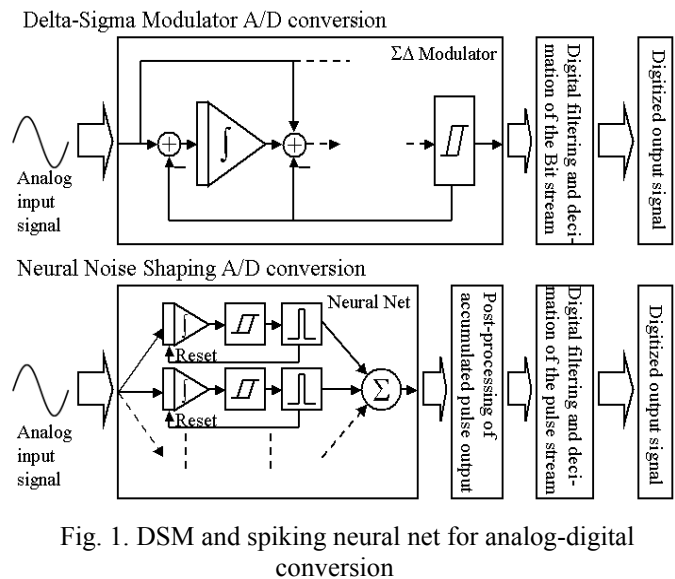

Fig. 1. DSM and spiking neural net for analog-digital conversion

The single loop structure of the DSM and the multiple modulator structure of the neural net is clearly evident. What has been omitted in the above illustration is the feedback of all pulsed outputs to all integrator inputs (Fig. 2), adding an additional degree of freedom to the Noise Shaping behavior when compared to simple parallel modulators, which would only contain a feedback to their own integrator. One important parameter defining the behaviour of these nets is the makeup of this feedback weight matrix (comparable to the coefficients of a DSM loop), which, as shown in [2] acts through inhibitory connections to decorrelate the individual neuron pulses, thus achieving a ratio of increase in SNR (Signal to Noise Ratio) to increase in the number of neurons well in excess of a simple parallel connection [8]. This behaviour is closer to the performance reported for parallel conventional DSMs in [7], where Galton et al. decorrelate input and output signals in a similar manner. In section II of this paper, we will attempt to show, that in analogy to DSM Theory mentioned above, the makeup of the weight matrix also influences the amount and size of the Noise Shaping exhibited by the network.

A second important aspect when comparing spiking neural networks to DSMs is the shape of the output signal, constituting the concatenated spikes for a neural net, and the state of the ADC for the DSM. In the former case, signal transmission with respect to absolute levels of noise and signal is very non-linear, depending on spike length, network frequency, variation of spiking frequency among single neurons, etc. Also, due to the unipolar nature of the spikes, there is always a varying DC offset present. In the latter case, the interplay between feedback coefficients and the bipolar nature of the ADC acts to very faithfully reproduce the input signal with respect to amplitude and offset, no spurious offset is introduced into the signal, and signal amplitude is the same as at the input.

Since spike generation is inherent to spiking neural nets, the mode of operation of the net can't be altered. However, to be able to apply Noise Shaping in spiking neural nets to real-world problems, the output signal has to closer reproduce the input signal. This can be achieved by post-processing of the output signal, as will be shown in section III. After post processing, the signal treatment would be essentially the same as in the DSM, with a digital filtering and decimation stage generating the high-resolution digital representation of the analog input signal [1] (Fig.1).

## II. SYSTEMATIC WEIGHT GENERATION

The weight matrices of spiking neural nets used for Noise Shaping can be optimized empirically, with good results. However, this necessitates a great deal of knowledge about both neural networks and DSMs, as well as being by its nature a time consuming process. This obstructs the technical use of neural networks as analog-digital converters and limits the number of network permutations that can be tested in a given time span. The goal has to be, then, to automate this weight generation process to increase its speed, improve performance, and make it independent of knowledge owned by the user.

### A. Motivation of Genetic Algorithms

The essential parameters to be optimized for a given network (network defined by topology and mean firing rate of single neurons) are the SNR for a given input signal and the spectral shape of the Noise Shaping (noise reduction in the baseband, no spikes or intermodulatory residues), so as to avoid detrimental network modes like oscillations. Because these performance characteristics can not be computed directly from the weight matrix, ordinary gradient-based optimization algorithms can not be used. Stochastic, unsupervised search methods like genetic algorithms are more suited to the task. Genetic algorithms perform their search through a mixture of stochastic search (mutation), different types of gradient search needing no information about the direction of the gradient (mating, crossover), and parallelism (large number of individuals, subpopulations). These algorithms are therefore ideally suited for optimizing an objective function which contains no direct relationship between network performance and network parameters and about whose hypersurface (shape and range) little a priori knowledge can be gathered [5,6]. For the simulations described herein, the Genetic Algorithm Toolbox for Matlab by Chipperfield et al. has been employed [5]. Individuals in the population utilized by the genetic algorithm were composed of a linear representation of the weight matrix. For a discussion of the effects of such a representation on the performance of the genetic algorithm, see Stanley at al. [9]. To give a short description of the GA employed for optimizing the network weights: A multi-population GA was utilized, containing 10 populations with 30 individuals, mutation rate of 0.004 per single gene, elitism activated, i.e. the top 15 percent parents are copied unchanged in each generation. Crossover is done on a gene by gene (i.e. weight by weight) basis, with a probability of 0.3. This probability is further equally subdivided into two modes of crossover. The first mode of crossover exchanges the gene (weight) between the two individuals, the second mode of crossover computes the average of the two genes and inserts this average in exchange for the original gene in both individuals. Individuals are selected for crossover, i.e. producing offspring, with stochastic universal sampling, meaning they are allowed to produce progressively more offspring based on their own fitness (value of the objective function). Migration between subpopulations is done every 20 generations on a ring structure basis, i.e. a subpopulation would exchange 15 percent randomly selected individuals with each of its two neighbors.

The algorithm starts out with randomly generated individuals, whose weight ranges are chosen to be between 0 and –0.2 because of theoretical considerations [2]. The fitness of each individual is computed according to (1) - (2) or (3). Individuals are selected for crossover based on their fitness, and the offspring produced by this crossover is subsequently mutated. The fitness of the offspring is computed, and the subpopulations are newly assembled from the best of the offspring and the best 15 percent of the parents. Every 20 generation, migration would be done at this point. The algorithm then loops back to the crossover selection, since the fitness values are already known, consisting of the old fitness values of the parents and the newly computed fitness values of the offspring. Usually, populations converged somewhere around the $400^{th}$ generation, so the algorithm was terminated after 500 generations. The longest run was carried out for 1200 generations, without appreciable further improvements.

The settings and algorithm for the GA have been adapted from a multi-population GA example contained in the toolbox. For further toolbox information, e.g. a description of stochastic universal sampling, see the documentation provided with [5].

The structure of the integrate-and-fire neural network is given in Fig. 2, it is composed of the spiking neurons, a weight matrix linking all pulse outputs with all integrator inputs, as well as a weight vector feeding the input signal to the neuron integrators. The neurons are ordinary non-leaky integrate-and-fire neurons, containing an integrator (receiving the summed input signal), comparator, and a pulse generator which produces the output pulse if the integrator has reached a certain threshold and which also resets the integrator. There is no refractoriness period implemented in this model, i.e. the integrator is open to input again immediately after being reset by the pulse. The combined output of the network is composed of a sum of all the pulse outputs.

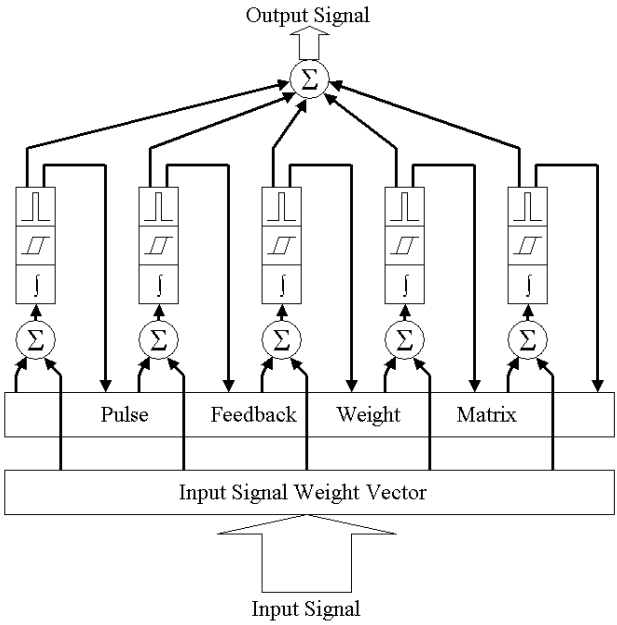

Fig. 2. Structure of the integrate-and-fire neural network

The weight matrices represented by the individuals were fed into this integrate-and-fire neural network in each generation, simulated, and the performance characteristics required by the objective function computed.

*B. Detrimental Effects*

The design of a proper objective function plays a large role in the performance of the genetic algorithm, i.e. its convergence and final result. As mentioned above, optimization criterions were the SNR and form of the Noise Shaping. Additionally, the mean firing rate of the neurons has to be constrained, since otherwise, the SNR would simply be increased by decreasing the average value of the weights, thereby increasing the mean firing rate and accordingly the OSR. Due to the constraints of the toolbox, there is only a single objective function, which is minimized by the algorithm. A possible objective function to achieve the above mentioned criteria is (1).

$$F_{Opt} = \frac{40dB}{SNR(dB) + 20dB} + \frac{\bar{f}_{Neuron} - 2kHz}{2kHz} + K_\sigma \quad (1)$$

With $K_\sigma$ defined as:

$$K_\sigma = \begin{cases} 2 * \left( \frac{\sigma_{f_{Neuron}}}{\bar{f}_{Neuron}} - 0.2 \right) & \text{for } \frac{\sigma_{f_{Neuron}}}{\bar{f}_{Neuron}} \geq 0.2 \\ 0 & \text{else} \end{cases} \quad (2)$$

The first addend acts to increase the SNR, the second to constrain the mean firing rate $\bar{f}_{Neuron}$ (averaged over all neurons) to 2 kHz, and the third acts to keep a check on the variations in firing frequency between the individual neurons, but only above a threshold as defined by $K_\sigma$. This ensures that there is a certain variation between neurons (to randomize the timing of individual pulses), but not too much (so that each neuron contributes about the same to the overall output signal). The factors weighting the addends were chosen in such a way that each of the above mentioned criteria has about the same contribution to the overall objective function, thus ensuring a balanced optimization process. The Decibel levels used with the SNR expect a SNR for final optimization of about 20-30 dB (specific for the chosen network topology and mean firing rate), this way also ensuring an equal weighting relative to the other addends. To ensure an equal frequency response over the whole base band, the properties (SNR, Noise Shaping) were computed with a random signal frequency for each individual in every generation.

However, this objective function did not achieve the desired results. The weight matrix generated by the genetic algorithm led, in all 20 trials of it, to a PLL (Phase locked Loop)-like mode of the network. This means that if there is no input signal present, the quantizer noise is approximately Noise-Shaped (Fig. 3, upper caption), with certain spurious elements, whereas in the presence of an input signal, the network locks onto the input signal and oscillates, thereby apparently increasing the SNR, but this SNR does not represent a true signal transmission characteristic of the network (Fig. 3, lower caption).

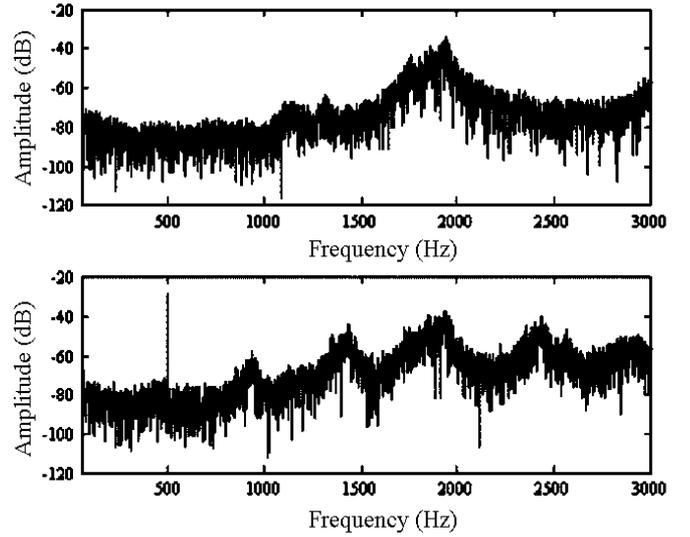

Fig. 3. Frequency spectrum of the summed output signal of the network, detrimental 'PLL-Mode' (Upper caption without input signal, lower with input signal present.)

This PLL-like mode can be seen in Fig. 3 from the sloping borders to the left and right of the input signal peak at 900 Hz. A true Noise Shaping behavior would only produce one peak directly at the signal frequency. The strong similarity between this diagram and the simulation results presented in [4] suggest that the network discussed in that paper has also entered a PLL-mode, thus the SNR results from the simulation reported by Marienborg et al. have to be questioned. This paper also offers an explanation for the apparent difference in results between simulation and hardware implementation in [4].

Apart from the difficult to formulate objective function and its effect on the performance of the algorithm, another detrimental effect is the stochastic variation in the performance characteristics due to the random initialization of the integrators contained in the neurons at each simulation. While these variations obviously affect the characteristics only to a slight degree, they do have impact on the final convergence of the genetic algorithm. As outlined in Fig. 4, at the start of the optimization (upper caption), the difference in performance between individuals is great enough to mask the stochastic variations due to initialization and a definite direction for optimization is present, but with almost converged populations (lower caption), the stochastic variations from generation to generation completely obscure the possible direction of optimization.

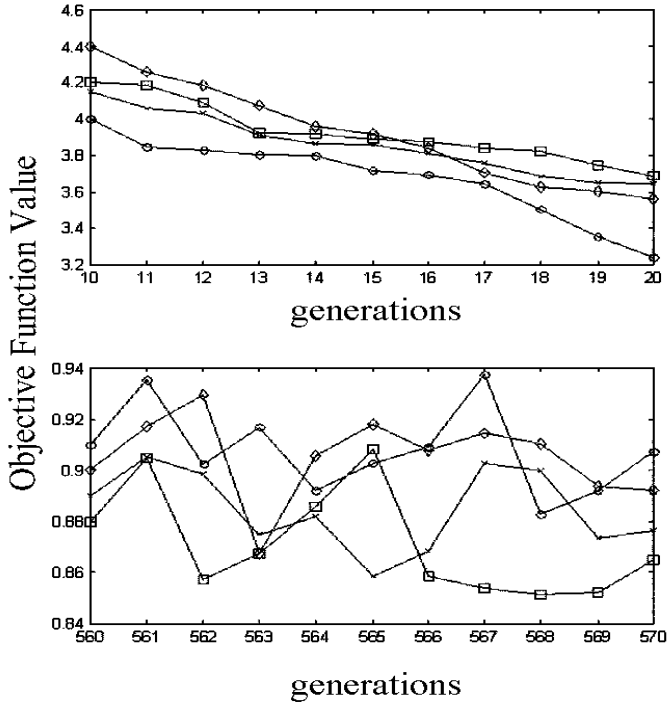

Fig. 4. Genetic Algorithm: Zoom on performance of first and last generations, displaying objective function value for best performing individual in each subpopulation.

This behavior has been partially counteracted by increasing the number of simulations in accordance with the progress of the optimization and averaging over these simulations, with the resultant penalty of a strong increase in computing time.

*C. Enhancement of Noise-Shaping Properties*

The objective function arrived at after numerous trials, which shows true Noise Shaping behavior and a definite increase in performance compared to randomly initialized weight matrices, is given in (3).

$$F_{Opt} = \frac{40dB}{SNR(dB)+20dB} + \frac{\bar{f}_{Neuron}-2kHz}{2kHz} + C * \sum_{i=1}^{N_{fmax}} \prod_{j=1}^{N_N} (e^{\frac{(f_i-\bar{f}_j)^2}{f_{norm}^2}} + C_{off}) \quad (3)$$

The first two addends are identical to (1), acting to increase SNR and constrain the average neuronal firing frequency to 2 kHz. The last addend envelops each individual mean neuronal firing frequency $\bar{f}_j$ with a bell-shaped curve of extension $f_{norm}$ and offset $C_{off}$ and takes the product of the value of the bell curve of all neurons ($N_N$ number of neurons), finally summing up this product for all baseband frequency bins $N_{fmax}$. C is a factor weighting this measure of frequency separation to ensure equal contribution to (3) relative to the other two addends. C is of course strongly dependent on the number of neurons, frequency bins, $f_{norm}$, and on $C_{off}$. It has to be chosen in such a way that a reasonably good frequency separation (as estimated from the presence/absence of spurious frequency spikes in the spectrum of the output signal) can be obtained, i.e. that the last addend has about a value of 1 for an optimized network. The principle behind this new measure of frequency separation for the individual neurons is illustrated in Fig. 5. $C_{off}$ can be used for adjusting the separation measure, low $C_{off}$ (e.g. 0.1) enforces more stringent separation of the individual frequencies. By adjusting $f_{norm}$, the distance between separated frequencies can be controlled.

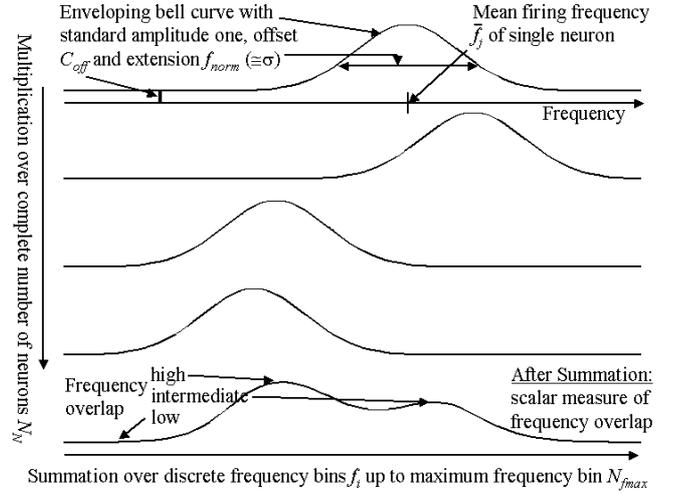

Fig. 5. 'Gaussian' Measure of frequency separation

The effect of this is to have a 'soft' measure of nearness/separation of the neuronal frequencies from each other, which is more effective in separating individual frequencies as a measure based on standard variation (2). The spectral behavior of two of these networks evolved for different mean neuronal firing frequencies is given in Fig. 6.

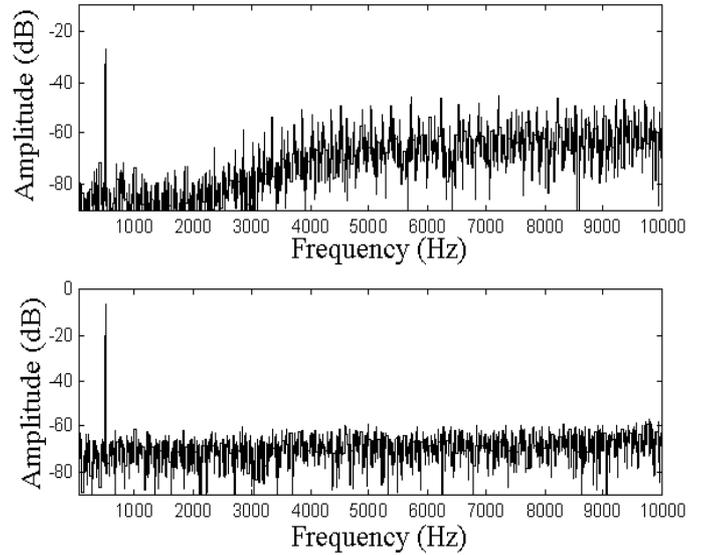

Fig. 6. Frequency spectrum of the summed output signal of the network, exhibiting true Noise Shaping behavior.

As can be seen from Fig. 6, the networks evolved according to (3) exhibit true Noise Shaping behavior, with a corresponding increase in noise attenuation when the network frequency is increased. Performance characteristics for the two networks are given in table 1.

TABLE I
PERFORMANCE CHARACTERISTICS OF THE FINAL NETWORK

|  | Network 1 | Network 2 |
|---|---|---|
| Mean firing frequency of individual neurons | 2,14 kHz | 20,3 kHz |
| Standard deviation | 0,83 kHz | 7,26 kHz |
| Oversampling ratio (OSR) | 10,7 | 101,5 |
| Signal to noise ratio (SNR) (10 Hz-1 kHz) | 22,1 dB | 52,9 dB |

Table 1 clearly shows that for a increase in mean firing frequency of factor 10, the corresponding SNR increases by more than 30 dB. This is well in excess of the performance reported in [2,4] and represents a definite advancement in employing spiking neural nets for Noise Shaping. Previously, authors have argued [2], that due to their one-integrator structure, neurons can not achieve Noise Shaping beyond first order (20 dB/dec OSR increase). However, by carefully and thoroughly optimizing the feedback loop (i.e. the inhibitory weight matrix), one can achieve a Noise Shaping behavior significantly higher than first order.

*D. Future Developments*

While a significant increase in the order of the Noise Shaping has been achieved beyond those previously reported [2,3,4], the objective functions used herein have only concerned themselves with optimizing OSR and the proper spectral operation of the network. Still, these objective functions have reached the limit of the genetic algorithms used for optimization [5]. To be able to use this network as part of an analog-digital converter on a VLSI-IC, there are a number of other directives one must optimize, e.g. sparsity of the connection matrix, emphasis on easy to implement local connections, ability to handle various mixtures of input signals or robustness to mismatch or statistical variations of the components on the IC.

A number of genetic algorithms have been developed for multiobjective function optimization [6], which could be tested and/or modified to extend the weight optimization to the above mentioned directives, computing the pareto optimal fronts of the various optimization criteria, along which the circuit designer can choose a suitable point with a corresponding tradeoff between these criteria.

Also, in this section, only the weight optimization of a given network topology has been researched. Topology and weight adjusting neural networks, as presented in [9], could offer a further increase in performance or at least applicability, with the circuit designer e.g. defining the width of the base band, the SNR, and the mean firing rate of the single neurons, and the genetic algorithm deciding on the number of neurons, network topology, and the connection weights.

Additionally, as can be seen from Fig. 6, the signal transmission in a spiking neural net is very non-linear, with an increase in OSR resulting not in a lowering of the noise floor (comparable to conventional DSMs), but in an increase of output signal amplitude, while the input signal to the network was the same for both simulations. Some of the Noise Shaping may be lost due to the noise floor caused by the random pulses. This behavior may also be influenced by the composition of the inhibitory weight matrix, and thus represents an additional optimization criterion.

## III. POST-PROCESSING OF SPIKING OUTPUT SIGNAL

As has been mentioned in the introduction, to be able to better handle the output signal of a spiking neural net, i.e. apply ordinary decimation filters, it has to undergo some form of post-processing [1]. This post-processing must achieve two main directives, minimize spurious DC-contributions and establish an improved relationship between signal amplitude at the input and output of the network, which is ideally independent of network parameters such as spike duration and mean firing frequency.

Previous implementations [4] simply computed the frequency spectrum of the output signal while only catching the leading edge of each pulse, thereby implicitly setting the pulse duration to the minimum time step of their measuring equipment. This avoids generating any valleys or peaks in the spectrum at the 1/T pseudofrequency (T being the pulse duration), but has the additional effect of making the output signal and noise amplitudes dependent on the minimum time step of the measuring equipment (Fig. 7).

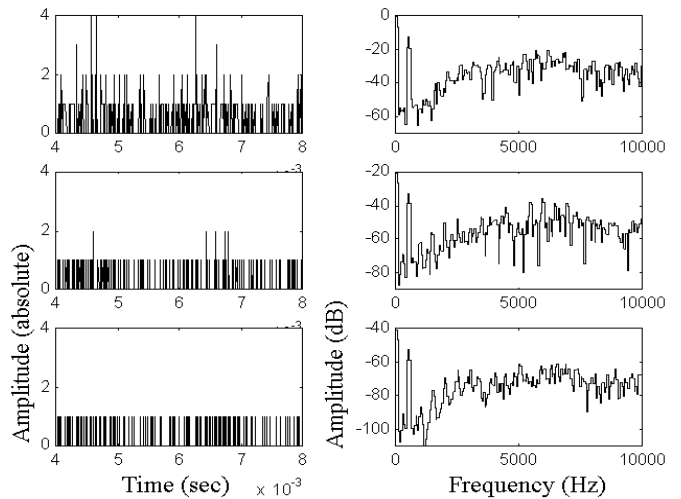

Fig. 7. Time base dependent amplitudes at the network output (from top to bottom, time bases of : 10µs, 1µs, 0.1µs, respectively).

In order to establish a simulation- or measurement-timebase-independent relationship between input and output signal, the individual pulse contributions to the output signal have to have a fixed duration. Concurrently, the pulse durations in the output signal have to be varied to avoid the above mentioned spurious effects at the 1/T pseudofrequency and harmonics. One possible way of achieving this is to convert the summed amplitude of the output spikes (upper caption Fig. 8) to a pulse duration signal, i.e. accumulating input spikes and releasing them at a constant rate. By doing this, a large number of amplitude levels (depending on the number of overlapping pulses) is transferred into a signal with two discrete amplitudes and a variable duration, coming closer to the PWM (Pulse Width Modulation)-signal of the output ADC in a DSM. This is done by applying the following three equations consecutively during each simulation time period.

$$A(t) = A(t-1) + \sum_{i=1}^{N_N} F_i(t) \qquad (4)$$

$$A(t) = \begin{cases} A(t)-1 & \text{for } A(t) \geq 1 \\ A(t) & \text{for } A(t) < 1 \end{cases} \qquad (5)$$

$$A_O(t) = \begin{cases} 1 & \text{for } A(t) \geq 1 \\ 0 & \text{for } A(t) < 1 \end{cases} \qquad (6)$$

In (4), all neuron output time functions $F_i(t)$ are summed and added to the accumulator state from the previous simulation time step *A(t-1)*. The neuron output time functions $F_i(t)$ are discrete-valued, 0 if no pulse is present, and 1 starting at the moment the integrator reaches the threshold and continuing for a set number of time steps, the fire pulse duration. Following this, in (5), the accumulator is decreased by a constant amount (in this case 1), until it reaches zero, with t being the number of discrete simulation time steps. The output signal of the accumulator $A_O(t)$ is given as the thresholded accumulator state $A(t)$ (6). The effect of this algorithm in shaping the output signal is illustrated in the lower caption of Fig. 8. To elaborate on this algorithm, take e.g. the pulse accumulation immediately following time step 6.15 ms in the upper caption of Fig. 8. The output of the accumulator $A_O(t)$ (lower caption Fig. 8) immediately switches to high and stays high to about 6.18 ms, when the accumulated pulses in $A(t)$ have been decreased enough by (5) to fall below the threshold expressed in (6).

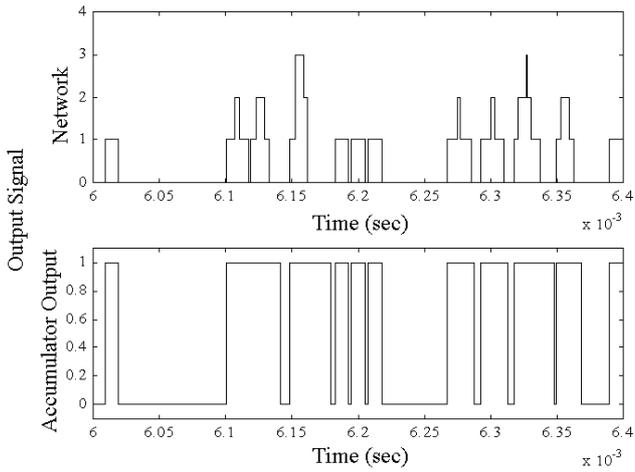

Fig. 8. Amplitude-summed output spikes and amplitude-duration converted network output signal.

By applying this algorithm, an increase in SNR of 2-3 dB for the given example can be achieved. This is mainly due to raising the output signal amplitude, thereby increasing distance to the background noise level of the network. However, the algorithm has not been fully successful in suppressing spurious effects at the 1/T pseudofrequencies (Fig. 9). This is caused by the fact that the algorithm acts primarily to increase pulse width, leaving the minimum pulse width unchanged if only a single pulse is present at a given time (e.g. time step 6.2 ms in Fig. 8). Pulses are only decreased in duration if they happen to overlap each other.

A new algorithm was developed that not only lengthens the duration of the pulses like the previous one, but also shortens them if they are situated in a time slot of very limited network output activity.

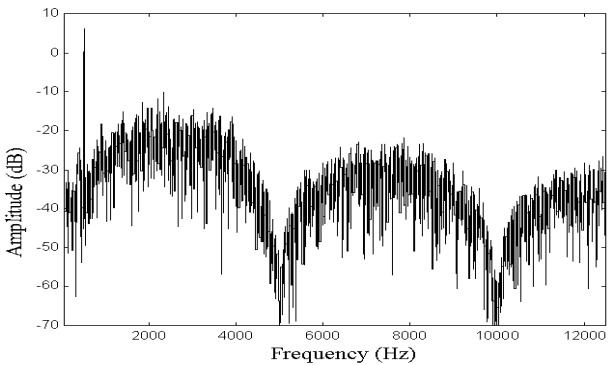

Fig. 9. Spurious effects in the network output spectrum at the 1/T pseudofrequencies due to the fixed minimum pulse width.

The following equation illustrates the new algorithm:

$$A(t) = \begin{cases} A(t-1) + u(t) - u(t + \varepsilon_{norm} * N_{fp}) \\ \quad for \quad \exists(Fi(t)) \geq 0 \\ A(t) \text{ else} \end{cases} \quad (7)$$

With $u(t)$ being the heaviside function, (7) has the effect of adding a variable pulse length to the accumulator, starting at the beginning of the original pulse $u(t)$ and ending at a time $u(t+\varepsilon_{norm}*N_{fp})$ dependent on the number of pulses at the starting time $N_{fp}$, normalized for best performance by the normalization factor $\varepsilon_{norm}$. The accumulator decrement and computation of the accumulator output $A_O(t)$ are still computed unchanged according to equations (5) and (6). The difference between equations (4) and (7) lies in the temporal extension of the factor added to the accumulator. In (4), the fixed duration of the fire pulse, represented by $F_i(t)$, is added to the accumulator, whereas in (7), if a fire pulse occurs, the current number of fire pulses $N_{fp}$ is computed and the increment of the accumulator is accordingly adjusted in duration, by increasing accumulator increment length with increasing number of pulses. This increase in duration is adjusted with $\varepsilon_{norm}$ so as to achieve the best SNR and similarity to conventional DSM output of the output signal, while preventing accumulator overflow. Basically, the equation acts to weight the pulse durations by a factor dependent on a moving average of the pulse intensities. By keeping the temporal length of the accumulator addend variable in this way, pseudo-frequency-effects as in Fig. 9 can be avoided. Fig. 10 illustrates the output behavior of the modified accumulator:

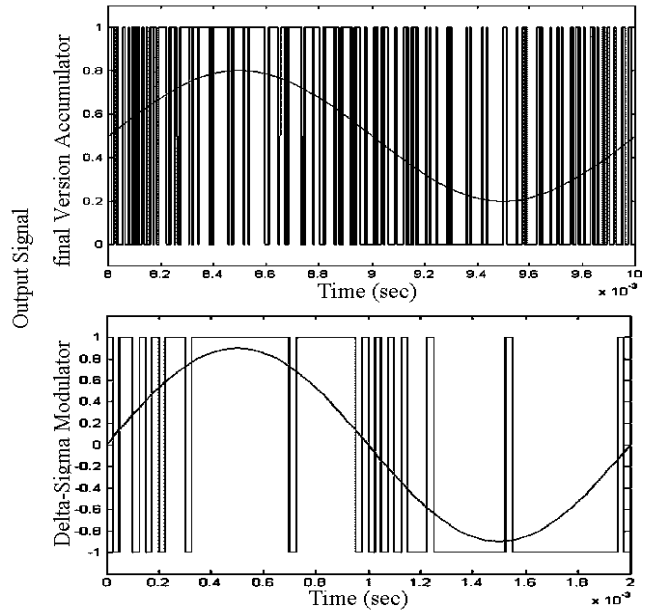

Fig. 10. Comparison of final duration-adjusted accumulator output signal and ADC-Output of conventional DSM with same input signal (sine wave)

In Fig 10, one can see a close resemblance between the accumulator output signal and the ADC-Output of a conventional DSM [1]. The unipolarity of the pulse output has been successfully converted to a bipolar pulse-duration signal, incorporating the same aspects of signal information as a conventional DSM and facilitating further signal processing according to intellectual property (IP) derived from DSM theory.

## IV. Conclusion

We have presented a scheme for automatic weight generation to be used in the application of networks of Integrate-and-Fire Neurons to Noise Shaping. As well, we have shown that appropriate post-processing of the output signal can improve output SNR, establish a better relationship between input and output signals, and align the network output signal with the ADC output of conventional DSMs. While the research discussed herein represents only a small step towards an eventual technical application of these networks as part of an oversampling analog-digital-converter, we have shown that reusing parts of the IP accumulated in DSM design, supplemented by appropriate newly developed algorithms, can significantly improve the overall performance of these nets.

Additional work is underway concerning several other aspects of neuronal Noise Shaping, e.g. various modifications in the feedback structure are being researched to significantly raise the order of noise shaping beyond the 30 dB/dec reported in this article, because this noise attenuation still falls far short of those reported for ordinary DSMs [1]. The eventual goal is some form of hybrid between DSMs and spiking neural nets, incorporating useful elements from both domains to form a new kind of analog-digital converter.

## Acknowledgment

The first author would like to gratefully acknowledge various discussions at the "Hochparallele VLSI-Systeme und Neuromikrelektronik" Chair (Technical University Dresden, Department of Electrical Engineering), as well as the financial support by the Sonderforschungsbereich 358, Teilprojekt A7 of the Deutsche Forschungsgemeinschaft. Many thanks also go to the two anonymous reviewers for their helpful comments on improving the quality and clarity of this paper.